\documentclass[aps,prl,reprint,groupedaddress,showpacs]{revtex4-1}

\usepackage{graphicx,amsfonts,latexsym,color,dcolumn,bm}
\usepackage{braket}
\usepackage{makeidx}
\usepackage{multirow}
\usepackage[dvipsnames,svgnames,table]{xcolor}
\usepackage{graphicx}
\usepackage{epstopdf}
\usepackage{ulem}
\usepackage{hyperref}
\usepackage{amsmath}
\usepackage{amssymb}

\usepackage{color}

\begin{document}

\newcommand{\be}{\begin{equation}}
\newcommand{\ee}{\end{equation}}
\newcommand{\bea}{\begin{eqnarray}}
\newcommand{\eea}{\end{eqnarray}}
\newcommand{\ad}{a^{\dag}}
\newcommand{\la}{\langle}
\newcommand{\ra}{\rangle}
\newcommand{\om}{\omega}

\title{Phase Selective Quantum Eraser}

\author{A. Heuer, G. Pieplow,  R. Menzel}
\affiliation{Photonik, Institut f\"ur Physik und Astronomie, Universit\"at Potsdam,  D-14469 Potsdam, Germany}

\date{\today}

\begin{abstract}

A quantum-eraser experiment is reported with photon pairs generated by two synchronously pumped parametric down converters coupled via induced coherence. 
The complementarity between which-source information and two-photon interference fringe visibility has been investigated by two coupled interferometers.
\end{abstract}

\pacs{42.50.Ar, 42.50.Dv}

\maketitle

The complementarity principle states that performing a measurement to expose both wave and particle aspect of a
single quantum object \cite{Bohr28} is not possible at a any given moment. The wave-like behavior often manifests itself in the appearance of interference fringes, while for example the '{\it welcher Weg}' information, a particle property, destroys the interference. 
Quantum-eraser experiments proved to be very useful to study the complementarity of wave and particle aspects of photons. 
Since the first proposal by Scully and Dr\"uhl \cite{Scully82}, quantum-eraser protocols have been discussed extensively, especially in the context of complementarity and fringe visibility \cite{Scully91,Walborn02, Kim00}. 
In optical interferometric studies \cite{Mandel95}, it was shown that interference results from the intrinsic indistinguishability of the photon's path. The fringe visibility vanishes, however , for distinguishable photon paths. 
Ultimately, this results in a direct relation of coherence and indistinguishability \cite{Mandel91}. 
Interference experiments that address this point using quantum eraser protocols can be found in \cite{Walborn02, Kim00}. 

In this work we present an experiment where we used two coupled parametric down conversion crystals in a cascaded arrangement. Wang, Zou and Mandel \cite{Zou91, Wang91} demonstrated that, in a setup such as that shown in Fig.1, coherence can be induced without induced emission, if the photon flux is so low that the probability for simultaneous generation of two photon pairs in the two crystals is negligible.
The coherence of the separately generated signal photons depends on the indistinguishability of the idler photons. For perfect matching of idler modes, the signal fields from BBO1 and BBO2 show perfect interference, while the interference is lost if the idler photon's origin becomes distinguishable. 
With this setup we can demonstrate the effect of phase memory with single photon interference \cite{Heuer14}. The phase memory is carried by the two photon state genearted by SPDC.  Recently the induced-coherence setup was used for a quantum imaging application \cite{Lemos14}.\\
In contrast to the last reference an additional interferometer for the idler photons has been added to our experiment to examine the interplay of indistinguishability and fringe visibility in two photon interference.  \\

\begin{figure}[t]
\includegraphics[width=8cm]{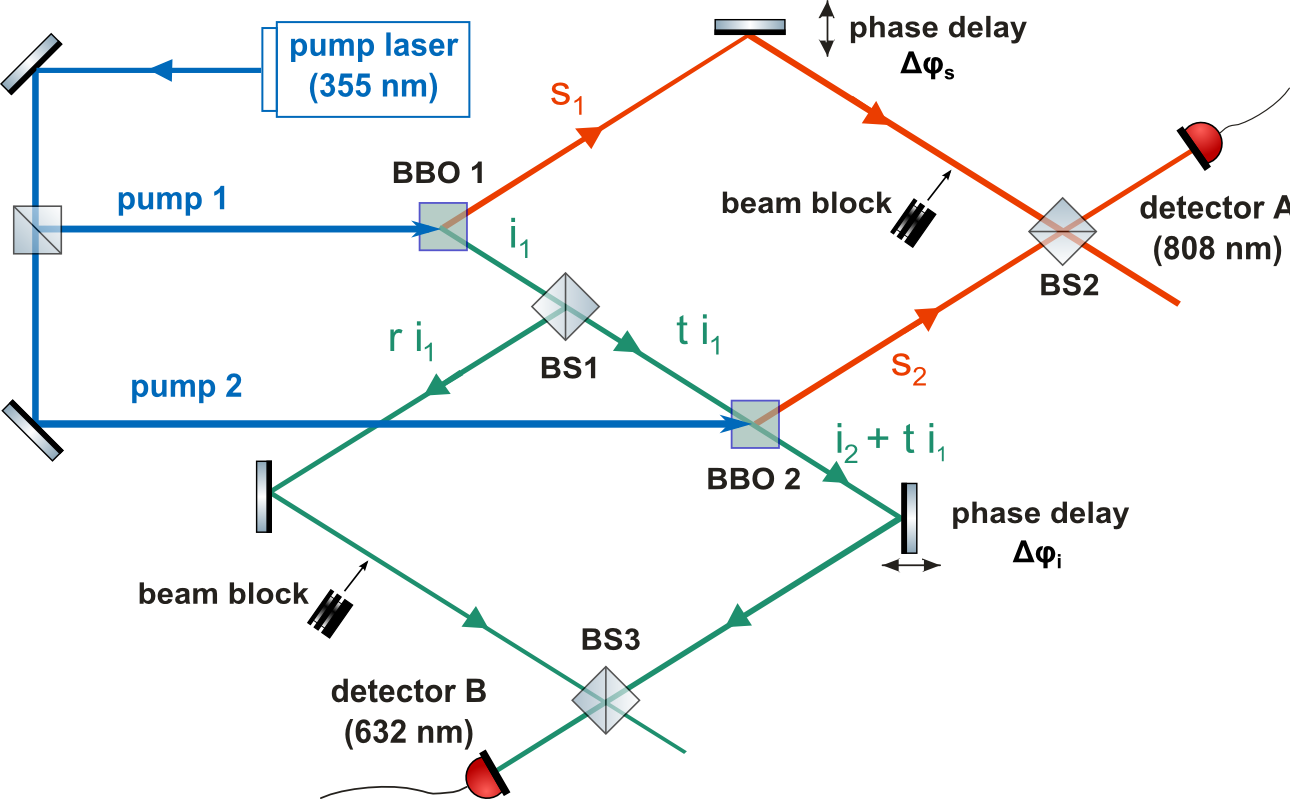}
\caption{Schematic of the experimental setup of two coupled interferometers based on induced coherence.}
\label{fig:experimentScheme}
\end{figure}

A schematic of our experimental setup is shown in Fig. 1. The pump light source is a frequency tripled optically pumped semiconductor laser continuously operating at a wavelength of 355 nm. A beam splitter was used to divide the pump beam into two almost identical coherent sub-beams to pump two BBO crystals as parametric down-converters. The crystals, with lengths of 4 mm each, were cut for degenerate collinear type I phase matching at 355 nm. In our case, the crystals were slightly tilted and the whole setup was arranged with filters and apertures to select signal photons at a wavelength of 808 nm and the corresponding idler photons at a wavelength of 632 nm.  
To achieve induced coherence without induced emission, the idler output i$_{1}$ of the first BBO1 crystal was matched perfectly with the idler output i$_{2}$ of the second crystal BBO2  (as described in \cite{Zou91}). A perfect overlap of the two beams is essential for a high visibility in the signal photon interference \cite{Grayson94b}. To observe the interference of the signal beams, both signal outputs of crystal BBO1 and crystal BBO2 were superimposed at the beam splitter BS2. To realize a variable phase delay, a high resolution delay line was introduced into the signal beam s$_{1}$. The minimum step size of the delay line was 20 nm. \\
An additional beam splitter BS1 in between both BBO crystals splits the idler beam i$_{1}$. It is recombined with the transmitted idler beam at the beam splitter  BS3, which effectively results in a Mach-Zehnder interferometer for the idler beam of crystal BBO1. The path length difference of the two beams was controlled using a variable trombone delay in the transmitted beam path.  
The photons were detected with fiber coupled avalanche photo diodes (SPCM-AQRH-13, Perkin Elmer) at the positions shown in Fig. 1. The measurement scenario can be altered with two beam blocks. To measure the classical induced coherence \cite{Zou91} the reflected idler beam of beam splitter BS1 can be blocked. For the coherence measurement of the idler beams a beam block can be inserted into the signal beam s$_{1}$.

In our first experimental run we examined the coherence properties of the signal beams. A beam block is used to block the reflected idler beam of beam splitter BS1. Thus the detector B will only see a fraction of the idler photons i$_{1}$ from crystal BBO$_{1}$ which are transmitted through the beam splitter BS1 and superimposed with the idler photons i$_{2}$ from crystal BBO$_{2}$. The pump power for the two crystals was 35 mW each. The resulting count rates at the detectors were 22,000 photons/sec and 14,000 photons/sec for  detector A and  B, respectively. Under these conditions the probability for simultaneous generation of two biphotons during the measuring interval of 2 ns is less than 10$^{-4}$. This insured induced coherence instead of stimulated emission as the mechanism for the generation of the second biphoton and thus, stimulated emission can be neglected. 

\begin{figure}[t]
\includegraphics[width=8.5 cm ]{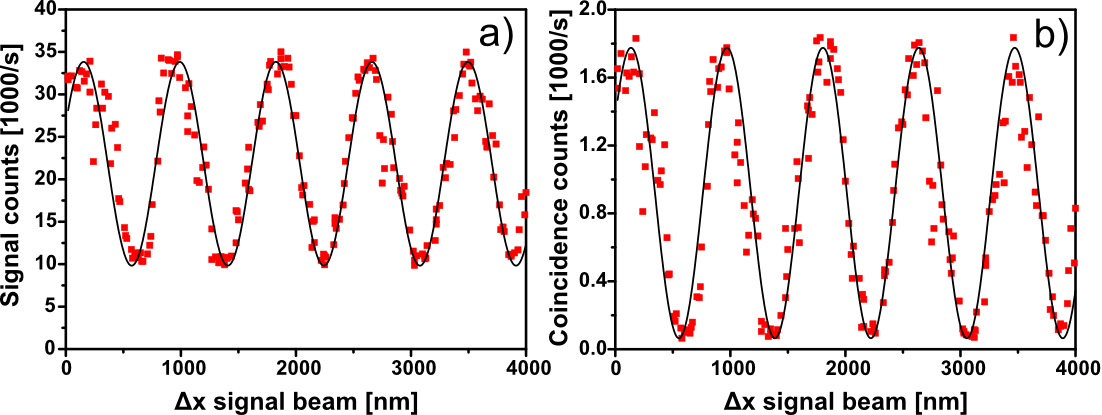}
\caption{Measured interference signal in a standard induced coherence setup (the reflected idler beam of  BS1 was blocked).  Both, the single photon rate of detector A (a) and the coincidence rate of detector A and detector B (b) were plotted as a function of the signal path length difference $\Delta$x$_{s}$. The solid curves represent the best sinusoidal fit. We found a period of 808 nm.} 
\end{figure}

The interference contrast is measured by varying the delay line in the signal beam s$_{1}$ and recording the count rates of signal detector A and the coincidence counts of detectors A and B. The results of the measurements corresponding to first-order interference are shown in Fig. 2(a) and those corresponding to second-order interference are showen in Fig. 2(b). The data was fitted by a least squares fitting routine. We found visibilities of 54\% for first-order interference and 91\% for second-order interference. These are remarkable values and indicate a high degree of induced coherence and good alignment of optical elements in the setup. In an induced coherence experiment, the visibility of first-order interference depends linearly on the transmissivity \cite{Wang91}, thus without the beam splitter a visibility of more than 80\% would have been recorded (amplitude transmissivity t = 0.67 of beam splitter BS1).

To adjust the alignment of the Mach-Zehnder interferometer for the idler photons, only the BBO1 crystal was pumped. A high contrast interference signal is recorded by detector B when the delay line of the upper idler beam is varied (see Fig. 3(a,b)). The visibilities of the fringes of the single count rate and the coincidence rate between detector B and A were both above 95\%.
If the second crystal is also pumped, the single photon count rate increases but the visibility is reduced significantly ($V \approx 50\%$ ) as seen in Fig. 3(c). This reduction cannot be attributed to an unbalanced interferometer. It seems that all the photons generated in the BBO2 crystal are added incoherently to the photons from the first crystal. This becomes even more apparent when the coincidence measurements (see Fig. 3(d)) are analyzed: Inserting a beam block in the signal arm s$_{1}$ and recording the coincidence rate between detector B and A does not show any variation, when modulating the path length difference of the idler beams.  From the quantum optical point of view it is clear that there is no coherence between the idler photons generated in BBO1 and BBO2 because of the non overlapping signal beams. The idler photon's pathways s$_{1}$ or s$_{2}$ are therefore totally distinguishable.

\begin{figure}[t]
\includegraphics[width=8.5 cm ]{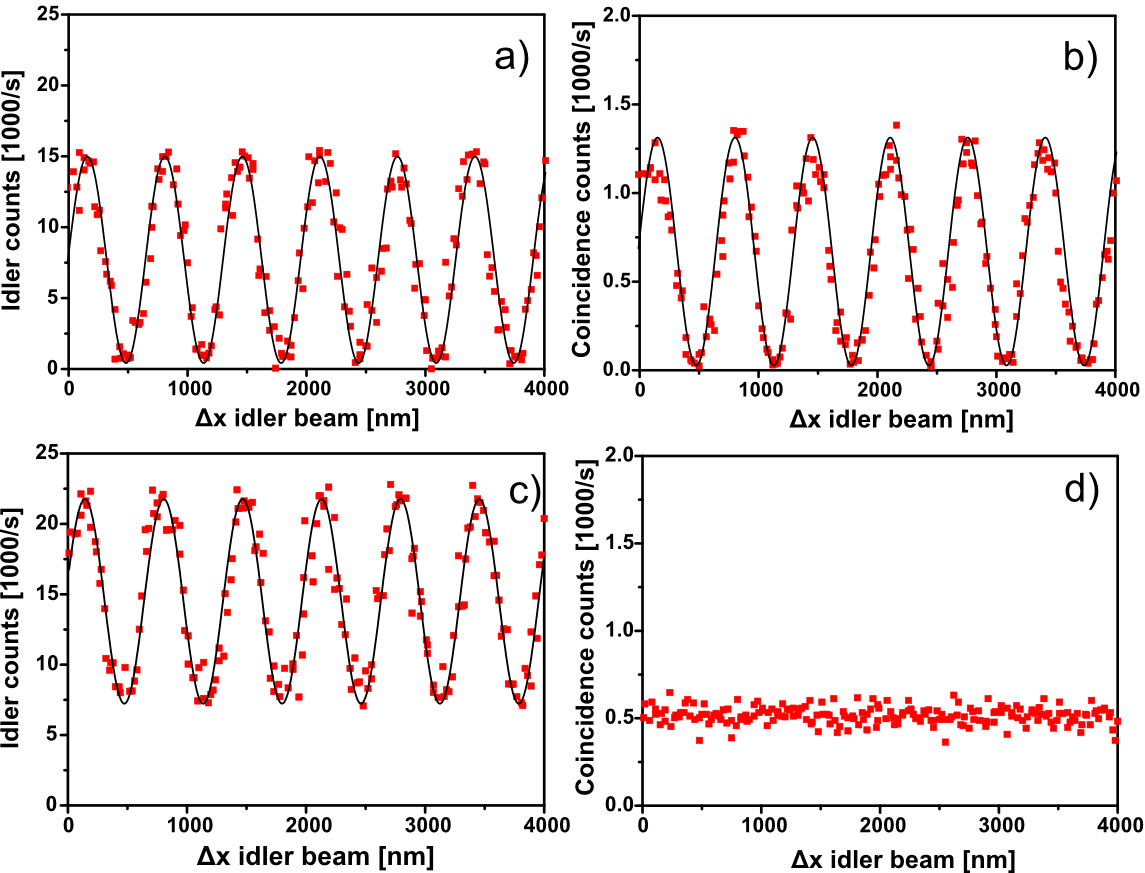}
\caption{Single rate (a,c) of detector B and coincidence rate (b,d) between detector A and  B as a function of the idler path length difference $\Delta$x$_{i}$. The solid curves represent the best fitt resulting in a period of 633 nm. During the measurement of (a,b) only crystal BBO1 was pumped. For the measurement of (c,d) both crystals were pumped and signal beam s$_{1}$ was blocked.}
\end{figure}

To erase the '{\it welcher Weg}' information of the idler photons, the beam block was removed. The signal photon detector A can can now measure photons from both crystals. This results in two coupled interferometers for the coincidence measurement of detectors A and B. The coincidence rate will be a function of both path length differences $\Delta$x$_{s}$ and $\Delta$x$_{i}$, respectively. In this run of the experiment we kept the phase difference of the idler beams   $\Delta\varphi_{i} = 2\pi/\lambda~\Delta x_{i}$ fixed.
Fig. 4(a) shows the measured coincidence rate as a function of the signal path length difference $\Delta$x$_{s}$ with an idler phase delay $\Delta\varphi_{i} \approx \pi/2$. 
The phase was determined by the amplitude of the single photon count rate of detector B. We expect the lowest count rate when there is no phase delay. This is due to the destructive interference of photons from the crystal BBO1. The count rate peaks at $\Delta\varphi_{i} = \pi$. The visibility of the interference signal is relatively high and reaches a value of 97\%. For $\Delta\varphi_{i} \approx 0$ the single count rate of detector B reaches a minimum and the visibility is reduced down to 40\% as can be seen in Fig.4(b).

\begin{figure}[t]
\includegraphics[width=8.5 cm ]{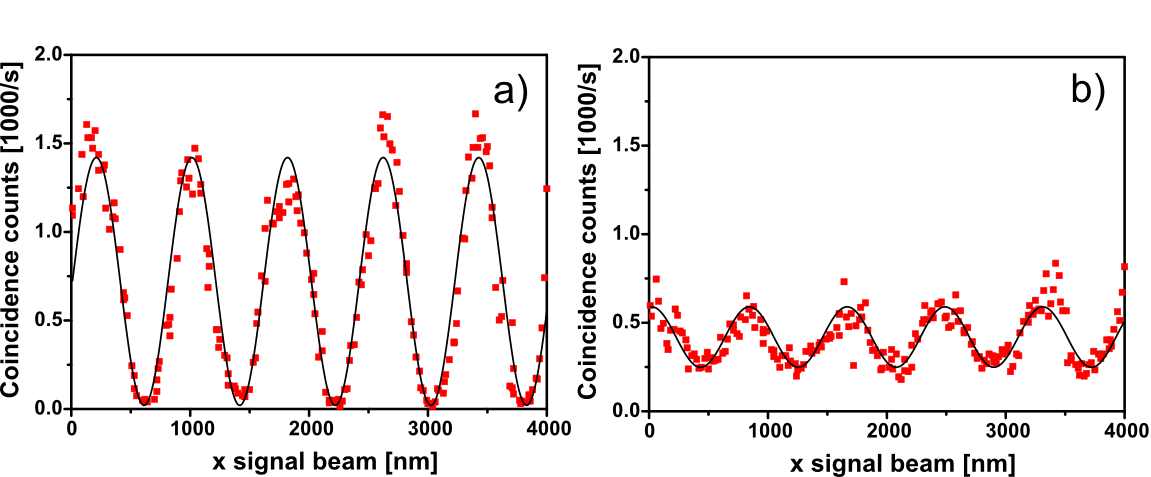}
\caption{Coincidence rate of detector A and B  depending on the signal path length difference $\Delta$x$_{s}$ for two different phase delays $\Delta \varphi_{i}$ of  the idler interferometer.
a) high interference contrast at a phase delay $\Delta \varphi_{i} \approx \pi/2 $ and b) low interference contrast at a phase delay $\Delta \varphi_{i} \approx 0 $. The solid curves represent the best sinusoidal fit. We found a period of 808 nm.}
\end{figure}

In the following section we discuss  results using a simplified theory \cite{Milonni96, Rehacek96, Wiseman00}. It captures the essence of what we measured without invoking a cumbersome formalism.  We can assume that the down-conversion efficiency $\gamma^2$ is so small that there is no more than a single photon pair present at any time. Due to this very low  pair generation rate, we assume that the pump fields p$_{1}$ and p$_{2}$ do not deplete and can hence be described with classical complex amplitudes A$_{p1}$ and A$_{p2}$. 
We work in the limit of perfect phase matching for all involved modes. We assume all fields to be monochromatic and that the interaction with the crystals is pointlike.  
With these assumptions, the field amplitudes for the signal/idler behind BBO1 are \cite{Milonni96}:
\begin{equation}
\begin{aligned}
A_{s1}= &~ a_{s01} +i ~C_{1} ~ a^{\dagger}_{i01}~,\\
A_{i1}= &~ a_{i01} + i ~C _{1}~ a^{\dagger}_{s01}~\\
\end{aligned}
\end{equation}
where a$_{s01}$ and a$_{i02}$ are the free, unperturbed operators satisfying the relation $a_{s01}\Ket{ \Psi} = a_{i01}\Ket{ \Psi} = 0$ for the initial state $\Ket{ \Psi}$. C$_{i}= \gamma A_{pi}$ is a constant that incorporates the crystal properties and the pump intesity. Due to the beam splitter BS1 (see Fig.~\ref{fig:experimentScheme}) the unperturbed idler operator a$_{i02}$ is subject to the usual transformation
\begin{equation}
\begin{aligned}
 a_{i02} = t_{1}~a_{i01} + r_{1}~a_{i0}~,
\label{eq:bs1tranfo}
\end{aligned}
\end{equation}
where t$_{1}$ and r$_{1}$ are the transmissivity and reflectivity of the beam splitter BS1 and a$_{i0}$ is the annihilation operator of the vacuum field mode in the empty port of the beam splitter. In our subsequent calculations we assume that the idler field i$_{02}$ that interacts with the pump p$_{2}$ in BBO2 is perfectly aligned with the transmitted idler field i$_{01}$ and that the conversion efficiency is so low that the generated idler field i$_{1}$ of BBO1 contributes negligibly (compared to the vacuum idler) to the generation of the signal s$_{2}$ in the second crystal.
After propagation through the setup (see Fig. 1) the total electric field amplitudes of the signal field at detector A and of the idler field at detector B are:
\begin{equation}
\begin{aligned}
A_{A}= &~ r_{2}~e^{i\Delta\varphi_s}\big(a_{s01}+iC_{1}  a^{\dagger}_{i01}\big)\\                    
&+t_{2}\big(a_{s02} +iC_{2} \big( t_1 a^{\dagger}_{i01}+ r_1a^{\dagger}_{i0} \big)\big)~,\\
A_{B}=&~r_{3}~r_{1}\big(a_{i01} + i C _{1}~a^{\dagger}_{s01}\big) + r_3 t_1 a_{i0}\\
&+t_3 e^{i\Delta\varphi_i}\big( t_1a_{i01}+ r_1a_{i0}+i t_1 C_1 a^{\dagger}_{s01}+i C_2 a^{\dagger}_{s02}\big)~,\\ 
\end{aligned}
\end{equation}
where $r_{x}$ and $t_{x}$ are the reflectivity and transmissivity of the respective beam splitter. The phase factors $\Delta\varphi_{i}$ and $\Delta\varphi_{s}$ are caused by the idler interferometer and the signal interferometer, respectively.
From $A_A$ and $A_B$ we can compute the normally ordered correlation functions that provide the counting rates measured by the ideal photon detectors A and B. 
Because signal and idler modes are initially unoccupied, we evaluate the correlation 
functions in their respective vacuum states.
We calculate the correlation functions to the first order of the $C$'s, which is consistent with our assumption that there is at most only one single biphoton in the apparatus during any detection interval. The signal photon counting rate at B is proportional to    
\begin{align}
&\begin{aligned}
					R_{B} \propto&~\la A_B^{\dagger}A_B\ra \\
          =&~|C_1\big(r_1 r_3+t_1 t_3 e^{i\Delta\varphi_i}\big)|^2\la a_{s01}\ad_{s01}\ra \\
          &+|C_2 t_3e^{i\Delta\varphi_i} |^2\la a_{s02}\ad_{s02}\ra
					\end{aligned}\\
&\hphantom{R_{B}}\, = ~|C_1\big(r_1 r_3+t_1 t_3 e^{i\Delta\varphi_i}\big)|^2 +|C_2 t_3|^2~. \label{eq:RBrate}
\end{align}
In accordance with the experimental findings (see Fig. 3(c)) equation \eqref{eq:RBrate} 
contains the incoherent contribution to the detector signal: The second term on the right of equation \eqref{eq:RBate} ($|C_2|^2$) corresponds to the rate of idler photon generation  in the BBO2 crystal. Although the indistinguishable idler photons induce the coherence of the signal photons there is no visible interference between the idler photons generated in BBO1 and in BBO2. 
However,
the joint detection rate of signal and idler photons shows high contrast second-order interference.
The coincidence counting rate $R_{AB}$ for detection events at detector A and detector B is 
\begin{equation}
\begin{aligned}
R_{AB}\propto&\la A_A^{\dagger}A_B^{\dagger}A_B A_A\ra \\
              =&~|C_1 r_2 e^{i\Delta\varphi_s}\big(r_1 r_3 + t_1 t_3 e^{i\Delta\varphi_i}\big) + C_2 t_2 t_3 e^{i\Delta\varphi_i} |^2~.
\end{aligned}\label{eq:pathinterference}
\end{equation}
The three terms in equation \eqref{eq:pathinterference} represent the three interfering paths that a signal and idler photon can take in order to be counted at A and at B simultaneously. The first probability amplitude 
(i) $i r_2 r_1 r_3 C_1e^{i\Delta\varphi_s}$ represents the following events: Signal photon s1 acquires the phase delay $\Delta\varphi_s$ and is reflected at $BS1$, while the idler i1 is reflected at BS1 and BS3.
(ii) $i r_2 t_1 t_3 C_1e^{i(\Delta\varphi_s + \Delta\varphi_i)}$ represents:
Signal photon s1 acquires the phase delay $\Delta\varphi_s$ and is reflected at BS2
while the idler photon i1 is transmitted trough BS1, aquires a phase delay $\Delta\varphi_i$
and is transmitted through BS3. 
(iii) $i t_2 t_3 C_2e^{i\Delta\varphi_i}$ represents: 
Signal photons are transmitted through BS2, the idler photon i2 acquires a phase delay $\Delta\varphi_i$, and is transmitted through BS2.

\begin{figure}[t]
\includegraphics[width=8.5 cm ]{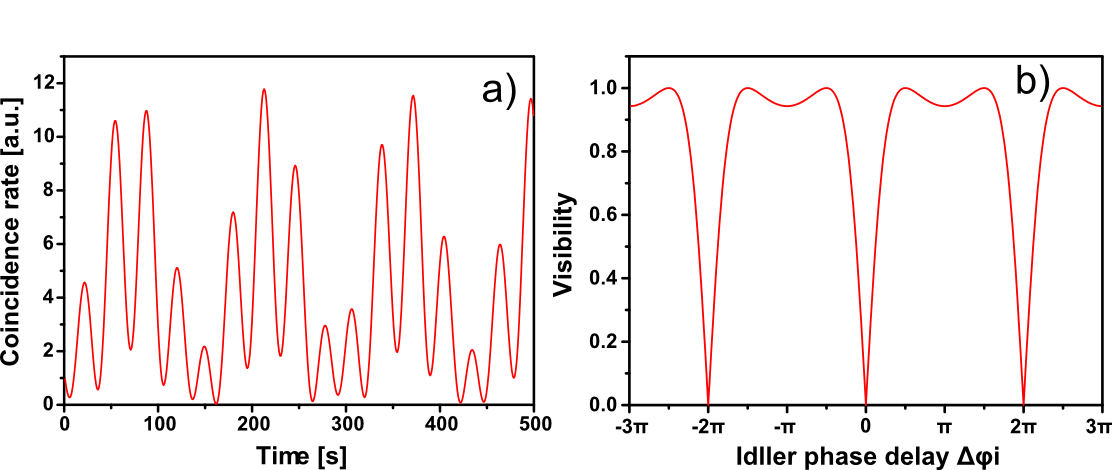}
\caption{a) Coincidence rate as a function of time: The path length differences of the signal and the idler interferometer $\Delta$x$_{s}$ and $\Delta$x$_{i}$ are varied at constant speed of 20~nm/s. b) Visibiltiy as a function of the idler phase delay  ${\Delta\varphi_i}$ as given by expression 7 }
\end{figure}

We now assume, for the sake of simplicity, equal conversion rates (C$_{1}=C_{2}$) and perfectly balanced beam splitters. The coincidence rate $R_{AB}$ is:
\begin{equation}
\begin{aligned}
R_{AB} = 2 -  cos(\Delta\varphi_i)-\sqrt{8}&\cos(\Delta\varphi_i/2)\times\\
&\sin(\Delta\varphi_s-\Delta\varphi_i/2)~.
\end{aligned}
\end{equation}
Since our experiment is based on two coupled interferometers, the coincidence rate varies with both phase delays. When both delays are varied simultaneously a beat structure manifests itself. Fig. 5(a) shows the calculated coincidence rate as a function of time. The path length differences of the signal and the idler interferometers $\Delta$x$_{s}$ and $\Delta$x$_{i}$ are varied at a constant speed of 20~nm/s. Although the path length differences are changed with the same velocity, the phase delay is different due to the different wavelengths of the signal and idler photons ($\lambda_s = 808~nm$ and $\lambda_i = 632~nm$). The beat structure indicates that there are certain values for $\Delta\varphi_s$ and $\Delta\varphi_i$ that maximize the fringe visiblity of the coincidence rate R$_{AB}$. If only the signal phase delay is varied, the expression for the fringe visibility as a function of the idler phase delay is:
\begin{equation}
\begin{aligned}
V(\Delta\varphi_i) = \frac {|\sqrt{8}cos(\Delta\varphi_i/2)|}{2 -  cos(\Delta\varphi_i)}~.
\end{aligned}
\end{equation}
The visibility is 0 for ${\Delta\varphi_i = n \cdot 2\pi}$ and is 1 for odd multiples of ${\pi/2}$ as can be seen in Fig 5(b). 
The high visibility close to 1 that we observed for ${\Delta\varphi_i = \pi/2}$ and a significantly reduced visibility  for  ${\Delta\varphi_i = 0}$ 
reproduces the theoretical findings [see Fig. (4)].
Due to the imperfect balance of the beam splitters BS1 and BS3, the contrast in Fig. (4) b) does not vanish.  
Instabilities in the setup which lead to fluctuations in ${\Delta\varphi_i}$ are another reason for not completely extinguishing the contrast. In combination with the very sharp resonance at ${\Delta\varphi_i = 0}$, measuring a visibility of 0 is very challenging.\\
In the ideal setup of perfectly balanced beam splitters one can explain the loss and recovery of interference contrast in a simple way: 
The idler photons originating from BBO1 will interfere destructively at the beam splitter BS$_3$ for a phase delay ${\Delta\varphi_i = n \cdot 2\pi}$. Thus, only the idler photons originating from the BBO2 crystal will be counted at detector B. This tags the signal photon's origin, which prevents interference in the signal interferometer. 
When balancing the rate of idler photons originating from BBO1 and BBO2 by adjusting the idler phase ${\Delta\varphi_i = \pi/2}$ the 'which crystal' tag of the signal photon is erased and interference in the signal/signal interferometer is restored. Our experiment is thus a realization of a quantum eraser. 
Intermediate states of the idler phase only make partial 'which crystal' information available and therefore they reduce the interference contrast. 
In summary, an experiment with two coupled interferometers for correlated signal and idler photons generated by spontaneous parametric down conversion has been presented. The coupling of the interferometers was achieved by induced coherence. The detuning of the idler interferometer allows for the selection of the origin of the photons. We thus demonstrated a quantum eraser protocol for the complementarity of wave-like and particle-like behavior of photons .

\section{Acknowledgment}
We are grateful to P. W. Milonni for valuable and illuminating discussions.

\end{document}